\documentclass[twocolumn,showpacs,preprintnumbers,amsmath,amssymb,prl,aps]{revtex4}

\usepackage{epsfig}% Include figure files
\usepackage{graphicx}% Include figure files
\usepackage{dcolumn}% Align table columns on decimal point
\usepackage{bm}% bold math

\begin{document}

\title{Fabrication of half metallicity in a ferromagnetic metal}

\author{Kalobaran Maiti}
 \altaffiliation{Electronic mail: kbmaiti@tifr.res.in}

\affiliation{Department of Condensed Matter Physics and Materials'
Science, Tata Institute of Fundamental Research, Homi Bhabha Road,
Colaba, Mumbai - 400 005, INDIA}

\date{\today}

\begin{abstract}

We investigate the growth of half metallic phase in a ferromagnetic
material using {\em state-of-the-art} full potential linearized
augmented plane wave method. To address the issue, we have
substituted Ti at the Ru-sites in SrRuO$_3$, where SrRuO$_3$ is a
ferromagnetic material. Calculated results establish Ti$^{4+}$
valence states (similar to SrTiO$_3$), which was predicted
experimentally. Thus, Ti substitution dilutes the Ru-O-Ru
connectivity, which is manifested in the calculated results in the
form of significant band narrowing leading to finite gap between
$t_{2g}$ and $e_g$ bands. At 75\% substitution, a large gap ($>$
2eV) appears at the Fermi level, $\epsilon_F$ in the up spin density
of states, while the down spin states contributes at $\epsilon_F$
characterizing the system a half-metallic ferromagnet. The $t_{2g} -
e_g$ gap can be tailored judiciously by tuning Ti concentrations to
minimize thermal effects, which is often the major bottleneck to
achieve high spin polarization at elevated temperatures in other
materials. This study, thus, provides a novel but simple way to
fabricate half-metallicity in ferromagnetic materials, which are
potential candidates for spin-based technology.

\end{abstract}


\maketitle

The search of half metallic ferromagnetic materials has seen an
explosive growth in the recent times due to its potential
technological applications. In these materials, the electronic
density of states (DOS) at the Fermi level, $\epsilon_F$ corresponds
to only one kind of spin, while the other spin density of states
exhibit an energy gap at $\epsilon_F$. Thus, in the polarized
condition, electronic conduction strongly depends on the spin of the
charge carriers; the material is insulating for one kind of spin and
metallic for the other. This unique property makes them ideal
candidates for the development of spin-based electronics. Various
theoretical studies predicted half metallicity in Heusler alloys
\cite{heusler}, double perovskites \cite{dd}, manganates
\cite{manganates}, CrO$_2$ \cite{cro2}, graphene nanoribbons
\cite{son} etc. However, experimental studies on very few materials
such as manganates \cite{manganates} and CrO$_2$ \cite{cro2}, etc.
exhibit half metallicity at low temperatures. Thermal fluctuations
often lead to a reduction in spin polarization at elevated
temperatures \cite{bluegel} making it difficult for technological
applications.

In this study, we investigate the evolution of the electronic
density of states in SrRu$_{1-x}$Ti$_x$O$_3$ as a function of $x$.
SrRuO$_3$ is a ferromagnetic metal with Curie temperature of 165 K.
Spin polarization at $\epsilon_F$ is found to be negative in the
ferromagnetic ground state \cite{kmband,spinpol}. SrTiO$_3$, on the
other hand, is a band insulator. Various experimental studies
\cite{jkim,ddti} suggest (4+) valence state of Ti in the
intermediate compositions (similar to SrTiO$_3$), which corresponds
to 3$d^0$ electronic configuration. Thus, in addition to disorder
effect, Ti substitution leads to a dilution of Ru-O-Ru connectivity.
Transport measurements in SrRu$_{1-x}$Ti$_x$O$_3$ exhibit a range of
novel phase transitions involving disorder induced correlated metal,
Anderson insulator, correlated insulator and band insulators
\cite{kkim} for different values of $x$.

\begin{figure}
\vspace{-14ex}
 \centerline{\epsfysize=3.5in \epsffile{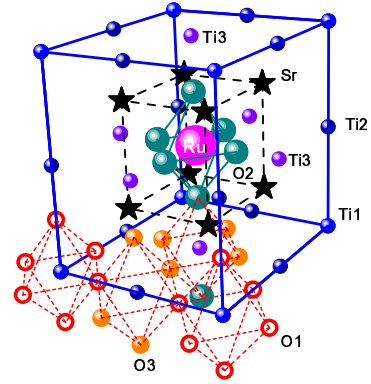}}
%\hspace{2ex}
 \vspace{-16ex}
\caption{(color online) Crystal structure of
SrRu$_{0.25}$Ti$_{0.75}$O$_3$. In order to obtain the structure of
SrRuTiO$_3$, we replaced Ti2 by Ru, and all the Ti and Ru sites are
made equivalent.}
 \vspace{-4ex}
\end{figure}

Using {\em ab initio} calculations, we find that Ti substitution at
Ru-sites in ferromagnetic SrRuO$_3$ leads to half metallicity. Here,
reduced Ru-O-Ru connectivity due to Ti-substitution leads to
significant narrowing of Ru 4$d$ band and thus, the up spin band
moves below $\epsilon_F$. Interestingly, the energy gap between
$t_{2g}$ and $e_g$ bands can be tuned by Ti-concentration. 75\%
substituted sample exhibits gap as high as 2 eV. Experimental
realization of such method on different systems would provide a new
direction in the search of HMFs for spin-based technology.

The electronic density of states of SrRu$_{1-x}$Ti$_x$O$_3$ for $x$
= 0.0, 0.5, 0.75 and 1.0 were calculated using state-of-the-art full
potential linearized augmented plane wave method (FLAPW) within the
local spin density approximations (LSDA) using {\scriptsize WIEN2K}
software \cite{wien}. The crystal structure of SrTiO$_3$ is cubic
with the lattice constant, $a$ = 3.905~\AA. SrRuO$_3$ possesses
close to cubic structure with small orthorhombic distortion. This is
manifested clearly by the similar density of states (DOS) of
SrRuO$_3$ in real structure vis-a-vis in the equivalent cubic
structure \cite{kmband}. Ti-substitution in SrRuO$_3$ leads the
system towards cubic structure. Thus, we have considered cubic
structure for all the calculations in this study. A typical unit
cell for SrRu$_{0.25}$Ti$_{0.75}$O$_3$ is shown in Fig.~1. There are
8 formula units in the unit cell constructed by doubling the lattice
constant of SrTiO$_3$. In order to preserve cubic symmetry, three
types of Ti are considered occupying corners (Ti1), edge centers
(Ti2) and face centered positions (Ti3). The body centered position
is occupied by Ru. There are three non-equivalent oxygens; O1 forms
the octahedra around Ti1-sites, O2 forms the octahedra around
Ru-sites and the rest of the oxygen positions are occupied by O3.
Thus, the connectivity between Ru-sites occurs via Ru-O2 bondings.
The muffin-tin radii ($R_{MT}$) for Sr, Ru, Ti and O were set to
1.16~\AA\, 0.95~\AA\, 0.95~\AA\, and 0.74~\AA\, respectively. The
convergence for different calculations were achieved considering 512
$k$ points within the first Brillouin zone. The error bar for the
energy convergence was set to $<$~0.25~meV per formula unit. In
every case, the charge convergence was achieved to be less than
10$^{-3}$ electronic charge.

\begin{figure}
\vspace{-2ex}
 \centerline{\epsfysize=4.0in \epsffile{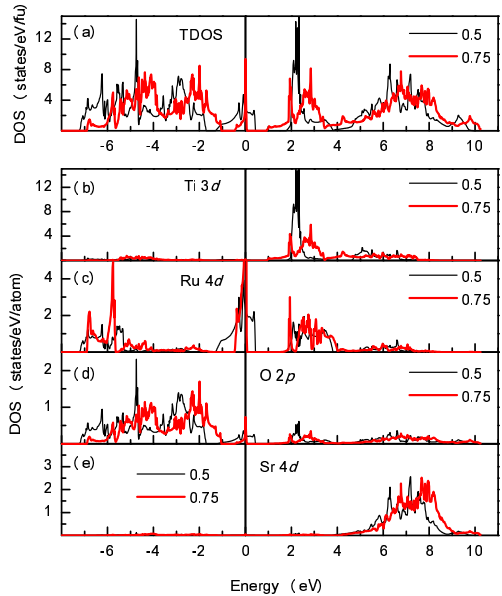}}
%\hspace{2ex}
 \vspace{-12ex}
\caption{(color online) (a) TDOS, (b) Ti 3$d$ PDOS, (c) Ru 4$d$
PDOS, (d) O 2$p$ PDOS and (e) Sr 4$d$ PDOS of
SrRu$_{1-x}$Ti$_x$O$_3$. Thin and thick solid lines represent DOS
corresponding to $x$ = 0.5 and 0.75, respectively.}
 \vspace{-4ex}
\end{figure}

In Fig.~2, We show the total DOS calculated for
SrRu$_{1-x}$Ti$_x$O$_3$ ($x$ = 0.5 and 0.75) and the partial DOS
obtained by projecting the eigenstates onto the Ti 3$d$, Ru 4$d$, O
2$p$ and Sr 4$d$ states. The figure exhibits 5 distinctly separable
features. The energy region -1.5 eV to -5 eV is primarily
contributed by O 2$p$ partial DOS with negligible contributions from
other electronic states. Thus, these contributions are characterized
due to the non-bonding O 2$p$ states. Sr 4$d$ partial DOS shown in
Fig. 2(e) appear above 5 eV. The peak appears to shift towards
higher energy with increasing $x$. This can be understood by
comparing the same in the end members, SrTiO$_3$ and SrRuO$_3$ as
demonstrated in Fig.~3. Sr 4$d$ states appear at much higher
energies in SrTiO$_3$ compared to that in SrRuO$_3$. One reason for
such a large shift may be related to the shift of the Fermi level to
the top of the O 2$p$ band in SrTiO$_3$. However, the shift of Sr
4$d$ band in the intermediate compositions, where the Fermi level is
pinned by the occupancy of the Ru 4$d$ band, indicates that the
Madelung potential at Sr-sites increases with the increase in Ti
concentrations.

Ti 3$d$ partial DOS appears 2 eV above the Fermi level. This clearly
demonstrates that the occupancy of Ti 3$d$ states is essentially
zero and hence correspond to Ti$^{4+}$ valency. Such valence states
was predicted in the x-ray photoemission spectra \cite{jkim}. This
study provides evidence of such effect theoretically within the
effective single particle approach itself. The width of the Ti 3$d$
$t_{2g}$ band is significantly small in $x$ = 0.5 sample ($\sim$
0.65 eV), which increases to 1.5 eV in $x$ = 0.75 sample and 2.5 eV
at $x$ = 1.0 (see Fig.~3).

\begin{figure}
\vspace{-2ex}
 \centerline{\epsfysize=4.0in \epsffile{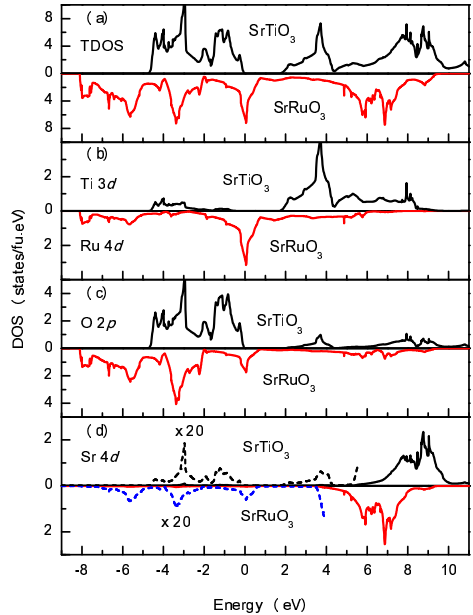}}
%\hspace{2ex}
 \vspace{-8ex}
\caption{(color online) (a) TDOS, (b) Ti 3$d$ and Ru 4$d$ PDOS, (c)
O 2$p$ PDOS and (d) Sr 4$d$ PDOS of SrTiO$_3$ and SrRuO$_3$. Dashed
line represent Sr 4$d$ PDOS rescaled by 20 times.}
 \vspace{-4ex}
\end{figure}

Ru 4$d$ partial DOS exhibit three regions. The narrow and intense
feature between the energy range -1.6 to 0.5 eV correspond to the
electronic states having $t_{2g}$ symmetry. The electronic states
above 1.8 eV appears due to Ru 4$d$ states having $e_g$ symmetry.
Notably, the O 2$p$ states also contribute in all the three energy
regions. Thus, DOS appearing below -5 eV can be attributed to the Ru
4$d$ - O 2$p$ bonding states having a large O 2$p$ character, and
the energy region above -1.5 eV are the anti-bonding states having
primarily Ru 4$d$ character. Most interestingly, both the compounds
exhibit metallic ground state. However, the $t_{2g}$ bandwidth, $W$
reduces significantly with the increase in $x$. While $W$ is close
to 2.6 eV in SrRuO$_3$, it is about 1.7 eV for $x$ = 0.5 and 0.54 eV
for $x$ = 0.75. Such reduction in $W$ is understandable as
Ti-substitution leads to a significant reduction in the hopping
interaction strength due to the reduced degree of Ru-O-Ru
connectivity. This is clearly evident in Fig.~1; if we assume
homogeneous distribution of Ru and Ti atoms in the solid, all the
RuO$_6$ octahedra are separated by TiO$_6$ octahedra at $x$ = 0.5.
At $x$ = 0.75, the number of Ru-[O-Ti-O]-Ru connectivity reduces to
half of that at $x$ = 0.5. Subsequently, $U/W$ ($U$ = local Coulomb
interactions strength) will increase significantly and presumably
play a role in the transport properties in these compositions
\cite{kkim}.

In order to understand the bonding of Ru 4$d$ electronic states with
various O 2$p$ states, we compare the Ru 4$d$ $t_{2g}$ and $e_g$
bands with the 2p bands corresponding to O1, O2 and O3 for $x$ =
0.75 and 0.5 sample in Fig. 4(a) and 4(b), respectively. All the
oxygens are equivalent in the $x$ = 0.5 sample. The energy
distribution of O2 2$p$ partial DOS is almost identical in Fig. 4(a)
to that observed in Ru 4$d$ partial DOS. This is expected as the
RuO$_6$ octahedra is formed by O2 atoms only. The width of the O2
2$p$ band is significantly larger than that of O1 and O3. The most
interesting observation is that the $t_{2g}$ and $e_g$ bands are
separated by a distinct energy gap. This gap is already visible in
Ru 4$d$ partial DOS of $x$ = 0.5 sample in Fig. 4(b) and is absent
in SrRuO$_3$ as shown in Fig.~3 and in the literature as well
\cite{kmband,djsingh}.

We calculate the crystal field splitting of the Ru 4$d$ band by
measuring the separation of the center of gravity of the Ru 4$d$
$t_{2g}$ and $e_g$ bands as shown in Fig. 4 by closed circles in
both the compositions. It is evident that crystal field splitting,
$\Delta$ remains almost the same ($\sim$ 2.1 eV) in both the
compositions and is very close to 2 eV found in SrRuO$_3$. Thus, the
large energy gap between the $t_{2g}$ and $e_g$ bands appears purely
due to the band narrowing. Such effect has strong implication in the
magnetic phase as described below.

\begin{figure}
\vspace{-2ex}
 \centerline{\epsfysize=4.0in \epsffile{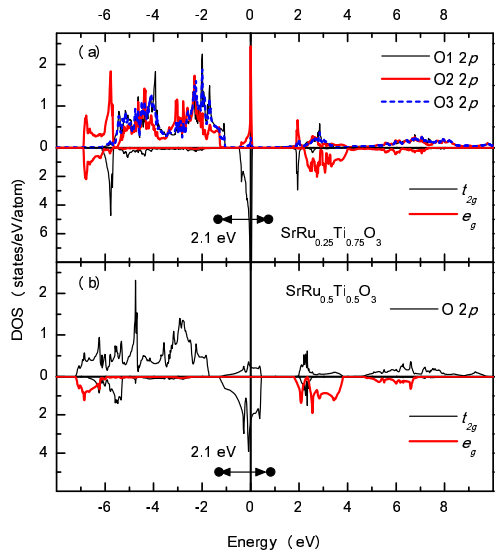}}
%\hspace{2ex}
 \vspace{-16ex}
\caption{(color online) Ru 4$d$ partial DOS with $t_{2g}$ and $e_g$
symmetry are compared with the O 2$p$ partial DOS in (a)
SrRu$_{0.25}$Ti$_{0.75}$O$_3$ and (b) SrRu$_{0.5}$Ti$_{0.5}$O$_3$.}
 \vspace{-4ex}
\end{figure}

It is already well established that the magnetic ground state can be
exactly described by these band structure calculations
\cite{kmband,hamada,ddprl,kmbairo3}. Thus, we have calculated the
ground state energies for ferromagnetic arrangement of moments of
the constituents using local spin-density approximations.
Interestingly, the eigen energy for the ferromagnetic ground state
in $x$ = 0.5 sample is 5.67 meV/fu lower than the lowest eigen
energy for the non-magnetic solution. This is higher than 1.2 meV/fu
observed in SrRuO$_3$ in real structure and significantly smaller
than 30.4 meV/fu observed in the equivalent cubic structure of
SrRuO$_3$. This energy difference between the non-magnetic and
magnetic solutions increases to 33.95 meV/fu in $x$ = 0.75. All
these results suggest that the stability of the ferromagnetic ground
state increases with the decrease in the degree of charge
delocalization of the valence electrons.

\begin{figure}
\vspace{-2ex}
 \centerline{\epsfysize=4.0in \epsffile{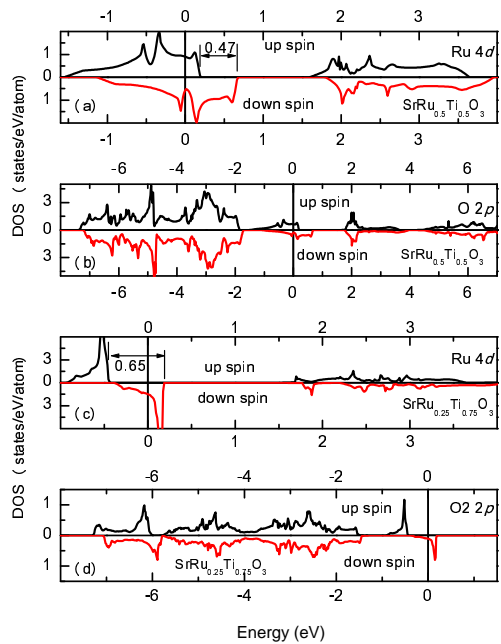}}
%\hspace{2ex}
 \vspace{-8ex}
\caption{(color online) Up and down spin density of stated
corresponding to (a) Ru 4$d$ in SrRu$_{0.5}$Ti$_{0.5}$O$_3$, (b) O
2$p$ in SrRu$_{0.5}$Ti$_{0.5}$O$_3$, (c) Ru 4$d$ in
SrRu$_{0.25}$Ti$_{0.75}$O$_3$, and (d) O 2$p$ in
SrRu$_{0.25}$Ti$_{0.75}$O$_3$. This figure demonstrates that band
narrowing in Ru 4$d$ band leads to a gap in the up spin channel
leading to half metallicity.}
 \vspace{-4ex}
\end{figure}

The spin magnetic moment centered at Ru-sites is found to be about
0.6~$\mu_B$ in $x$ = 0.5 sample. Interestingly, magnetic moment at
the interstitial electronic states is significantly large ($\sim$
0.36 ~$\mu_B$). The moment at the O sites is about 0.05 ~$\mu_B$.
The Ti sites also exhibit very small moment ($\sim$ -0.03 ~$\mu_B$).
Thus the total magnetic moment of the solid becomes 1.24 ~$\mu_B$
per Ru-atom. This is very similar to that observed (1.2 ~$\mu_B$) in
SrRuO$_3$. The magnetic moments increase significantly with the
increase in $x$. The moments at Ru site becomes 0.88 ~$\mu_B$ in $x$
= 0.75 sample. The moments of the interstitial states and 2$p$
states at O2 sites also enhance to 0.66 ~$\mu_B$ and 0.066 ~$\mu_B$,
respectively. Thus, the total moment turns out to be 1.99 ~$\mu_B$,
which is very close to the spin only value of 2 ~$\mu_B$
corresponding to Ru 4$t_{2g}^4$ electronic configuration. It is to
note here that although the local moment of the highly extended 4$d$
states is significantly smaller than the spin only value as opposed
to the case in 3$d$ transition metal oxides \cite{ddprl}, Ru 4$d$
moment induces a large degree of polarization in the interstitial
and O 2$p$ electrons. These results evidently suggest applicability
of Stoner description to capture magnetic properties of these
systems.

In order to investigate the exchange splitting and the character of
density of states in the vicinity of $\epsilon_F$, we plot the
spin-resolved DOS corresponding to Ru 4$d$ and O 2$p$ partial DOS in
Fig.~5. In the $x$ = 0.5 sample, both the up and down spin states
contribute at $\epsilon_F$ and the exchange splitting is found to be
about 0.47 eV. This is again very similar to the case in SrRuO$_3$
\cite{kmband}. The exchange splitting increases to 0.65 eV in $x$ =
0.75 sample as shown in the figure. Interestingly, the up spin band
moves significantly below $\epsilon_F$ and the contributions at
$\epsilon_F$ appears only due to the down spin states indicating a
half-metallic behavior. No contribution of the up spin states
observed in the total density of states (not shown here).
Considering the paucity of half-metallic materials for various
technological applications, achieving half metallicity in the
ferromagnetic SrRuO$_3$ by Ti-substitution is remarkable.

It is believed that the half metallicity can be achieved via strong
$d-d$ hybridization in Heusler alloys involving two transition metal
elements in the compound \cite{galanakis}. In transition metal
oxides, often doping of large amount of electrons or holes leads to
a shift of the Fermi level towards the energy gap of one spin
channel leading to half metallicity \cite{manganates}. The primary
difficulty to use these systems in technological applications is the
loss of half metallicity at elevated temperatures, where thermal
excitations leads to significant mixing of various spin channels due
to small energy gap at $\epsilon_F$ \cite{bluegel}. In the present
case, mechanism to achieve half metallicity is simple and easily
achievable experimentally. The most important aspect is that the
energy gap between $t_{2g}$ and $e_g$ bands can be tailored
judiciously by tuning the composition to minimize thermal effects.

In summary, we investigate the possibility of fabricating half
metallicity by Ti-substitution at the Ru-sites in a ferromagnetic
material, SrRuO$_3$. The calculated results using FLAPW method
within the local spin density approximations reveal tetravalency of
Ti in all the compositions consistent with the experimental
predictions. The Ru 4$d$ band exhibit significant narrowing with the
increase in Ti-substitution; the crystal field splitting remains
almost the same across the whole series. Thus, an energy gap
develops between the $t_{2g}$ and $e_g$ bands, which gradually grows
with the increase in $x$. Consequently, the up spin density of
states exhibit an energy gap at the Fermi level, while the down spin
states still contribute leading to half metallicity. Most
interestingly, the $t_{2g} - e_g$ gap can be engineered by tuning
$x$ and thus spin mixing effects due to thermal excitations can be
minimized. This study thus provide a novel but simple way to
fabricate half metallicity in ferromagnetic materials, which are
potential candidates for spin based technology. Experimental
realization of this method would help both chemists and physicists
to cultivate new materials. In addition, this study demonstrates
that effective single particle approaches provide a remarkable
description of the electronic properties of these systems, which are
predicted experimentally.

%\pagebreak

\end{document}